\documentclass[aps, prl, twocolumn, floatfix]{revtex4}
\usepackage{epsfig,amsmath,amsthm,amssymb}
\usepackage{./mciteplus}

\begin{document}

\title{Prospects for identifying the sources of the Galactic cosmic rays with IceCube}

\author{Francis Halzen, Alexander Kappes\footnote{on leave of absence
from Universit\"at Erlangen-N\"urnberg, Physikalisches Institut,
D-91058, Erlangen, Germany} and Aongus \'O\,Murchadha}

\affiliation{Department of Physics, University of Wisconsin,
Madison, Wisconsin 53706, USA}

\begin{abstract}We quantitatively address whether IceCube, a
kilometer-scale neutrino detector under construction at the South
Pole, can observe neutrinos pointing back at the accelerators of the
Galactic cosmic rays. The photon flux from candidate sources
identified by the Milagro detector in a survey of the TeV sky is
consistent with the flux expected from a typical cosmic-ray generating
supernova remnant interacting with the interstellar medium. We show
here that IceCube can provide incontrovertible evidence of cosmic-ray
acceleration in these sources by detecting neutrinos. We find that the
signal is optimally identified by specializing to events with energies
above 30\,TeV where the atmospheric neutrino background is low. We
conclude that evidence for a correlation between the Milagro and
IceCube sky maps should be conclusive after several years.
\end{abstract}

\maketitle

\section{Gamma-ray observations}
It is believed that Galactic accelerators are powered by the
conversion of $10^{50}$\,erg of energy into particle acceleration by
diffusive shocks associated with young (1,000--10,000 year old)
supernova remnants expanding into the interstellar medium
\cite{aa:75:386}. The cosmic rays will interact with atoms in the
interstellar medium to produce pions that decay into photons and
neutrinos. Dense molecular clouds, often found in star-forming regions
where the supernovae explode, are particularly efficient at converting
protons into pions that decay into ``pionic'' gamma rays and neutrinos.
These provide us with indirect but additional evidence for cosmic-ray
acceleration and, unlike the remnants seen alone, there is no
electromagnetic contribution to the TeV radiation that is difficult to
differentiate from the pionic gamma rays. The existence of the ``knee''
tells us that there must exist Galactic cosmic-ray sources producing
protons with energies of several PeV. These ``Pevatrons'' will produce
pionic gamma rays whose spectrum extends to several hundred TeV
without cut-off in interactions with the interstellar medium, in
particular with dense molecular clouds. By straightforward energetics
arguments such sources must emerge in global sky surveys with the
sensitivity of the Milagro experiment \cite{apj:658:L33}. We will
argue that one Pevatron, MGRO\,J1908+06, has likely been identified
among five candidates in the current sky map.

Supernovae associated with molecular clouds are a common feature of
associations of thousands of OB stars that exist throughout the
Galactic plane. Some of the first resolved images of sources in TeV
gamma rays were of the supernova remnants RX\,J1713.7-3946
\cite{na:432:75,aa:433:229} and HESS\,J1745-290 which illuminate
nearby molecular clouds to produce a signal of TeV gamma rays
\cite{na:439:695}. Although not visible to H.E.S.S., possible evidence has been
accumulating for the production of cosmic rays in the Cygnus region of
the Galactic plane from a variety of experiments
\cite{aa:393:L37,aa:423:415,apj:658:1062,astro-ph:0611731,apj:658:L33}.
Most intriguing is a Milagro report of an excess of events from the
Cygnus region at the $10.9\,\sigma$ level \cite{apj:658:L33}. The
observed flux within a $3^\circ \times 3^\circ$ window is 70\% of the
Crab at the median detected energy of 12~TeV and is centered on a
source previously sighted by HEGRA. Such a flux largely exceeds the
one reported by the HEGRA Collaboration, implying that there could be
a population of unresolved TeV $\gamma$-ray sources within the Cygnus
OB2 association. In fact, they report a hotspot, christened
MGRO\,J2019+37 \cite{apj:658:L33}. A fit to a circular two-dimensional
Gaussian yields a width of $\sigma=(0.32 \pm 0.12)^\circ$, which for a
distance of 1.7\,kpc suggests a source radius of about 9\,pc. As the
brightest hotspot in the Milagro map of the Cygnus region, it
represents a flux of 0.5 Crab above 12.5\,TeV.

To date, the Milagro collaboration has identified eight Galactic
sources of high-energy gamma rays. On the basis of prior observations
some of these sources appear to correspond to objects unlikely to be
significant sources of the Galactic cosmic rays. For example, three
Milagro hotspots are at the same locations as the Crab nebula,
Geminga, and the Boomerang nebula. As these objects are known to be
pulsar-wind nebulae, and therefore not likely to be significant proton
accelerators, we do not consider them in the context of this study.
Three of these sources, MGRO\,J1908+06, MGRO\,J2019+37, and
MGRO\,J2031+41, have post-trial significances of $\ge4.9\,\sigma$
\cite{apj:664:L91} (the only other Milagro source of such statistical
significance being the Crab nebula). The remaining two
hotspots---candidate sources C1 (MGRO\,J2043+36) and C2
(MGRO\,J2032+37)---are located within the Cygnus region of the Galaxy
at Galactic longitudes of $77^{\circ}$ and $76^{\circ}$,
respectively. Another potential hotspot, MGRO\,J1852+01, falls
currently somewhat below the threshold set by the Milagro
Collaboration for candidate sources. If confirmed it will be the
strongest source in Milagro's entire sky map with a flux about 2.5
times higher than MGRO\,J2019+37
\cite{talk:gsfc2007:abdo}. In the analysis that follows, we will
consider the five identified Milagro hotspots as our candidate
cosmic-ray accelerators and evaluate the impact of MGRO\,J1852+01 in
the event that it is later confirmed as a source.

We focus in particular on MGRO\,J1908+06. The H.E.S.S.\ observations
of this source reveal a spectrum consistent with a $E^{-2}$ dependence
from 400\,GeV to 40\,TeV without evidence for a cut-off
\cite{proc:icrc07:djannati-atai:1}. In a follow-up analysis
\cite{talk:heidelberg2008:casanova} the Milagro Collaboration showed that its own data
are consistent with an extension of the H.E.S.S.\ spectrum to at least
90\,TeV (Fig.~\ref{fig1}). This is suggestive of pionic gamma rays
from a Pevatron whose cosmic-ray beam extends to the `knee' in the
cosmic-ray spectrum at PeV energies. Another source with a measured
spectrum consistent with $E^{-2}$ is MGRO\,J2031+41
\cite{arxiv:0801.2391}. The lower flux measured by MAGIC can be
attributed to the problem of background estimation for Cherenkov
telescopes in a high density environment like the Cygnus region.

Not all the sources have known lower-energy counterparts,
however. Although the H.E.S.S.\ telescope array discovered a GeV-TeV
counterpart to MGRO\,J1908+06 and MAGIC to MGRO\,J2031+41, the VERITAS
telescopes failed to detect an excess at the location of
MGRO\,J2019+37 \cite{proc:icrc07:Kieda:1}. A possible reason for this
distinction is that this source, located in the Cygnus region of the
Galaxy, may not be the accelerator but a nearby molecular cloud
illuminated by a Pevatron beam. While the pionic gamma ray spectrum
extends to hundreds of TeV, it is expected to be suppressed in the TeV
search window of VERITAS \cite{apjl:665:l131}. Indeed, there could
be many potential accelerators in the Cygnus region, one of the
principal star-forming areas of the Galaxy.

In conclusion, evidence tracing the production of these or any other
sources of TeV gamma rays to pions produced by cosmic-ray accelerators
has been elusive. It is one of the main missions of neutrino
telescopes to produce incontrovertible evidence for cosmic-ray
production by detecting neutrinos associated with the
sources. Particle physics is sufficient to compute the neutrino fluxes
associated with the sources discussed. We evaluate in detail the
sensitivity of IceCube, the first kilometer-scale neutrino observatory
now half complete, to the Milagro sources assuming that they represent
the imprint of the Galactic cosmic-ray accelerators on the TeV
sky. Here, we include for the first time in these kind of calculations
the effect of a finite energy resolution of the detector and a
zenith-angle dependent angular resolution. While the number of events
with energies of tens of TeV is relatively low, we establish that this
is optimally the energy region where the atmospheric neutrino
background is suppressed and an excess from these sources can be
statistically established. While observing individual sources may in
some cases be challenging, we conclude that evidence for a correlation
between the Milagro and IceCube sky maps should be conclusive after
several years.

It is important to emphasize that the photon flux from the Milagro
sources is consistent with the flux expected from a typical cosmic ray
generating supernova remnant interacting with the interstellar medium
(see for instance \cite{aa:75:386}). In other words, the TeV flux is
consistent with the energetics that are required to power the
cosmic-ray flux in the Galaxy. Alternative candidates such as
micro-quasars have been suggested for the sources of the Galactic
cosmic rays. If that were the case, cosmic-ray energetics would
require that they leave their imprint on the Milagro sky map, but none
have so far been observed.
 
\section{Neutrinos from Gamma-ray sources}
\begin{figure}
\epsfig{file=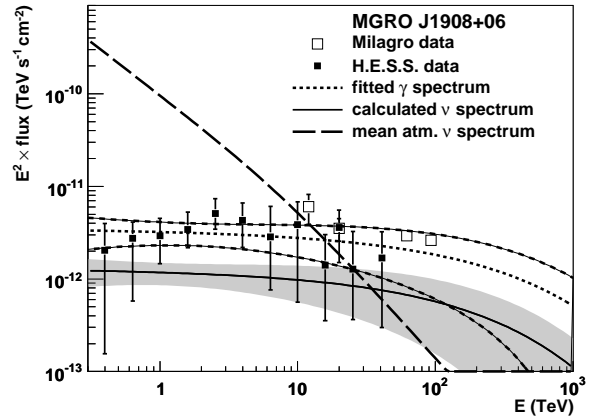, scale=0.43}
\caption{The $\gamma$-ray and neutrino fluxes from MGRO\,J1908+06. The
hollow/shaded regions surrounding the fluxes represent the range in the
spectra due to statistical and systematic uncertainties. Also shown is
the flux of atmospheric neutrinos at the same zenith angle as the
source (dashed line), taking into account the source size and angular
resolution.}
\label{fig1}
\end{figure}

\begin{figure}
\epsfig{file=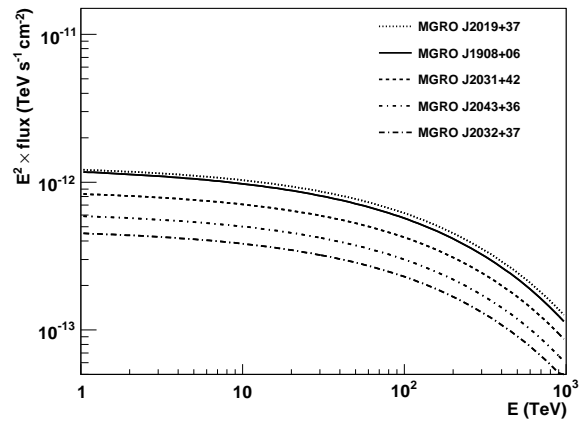, scale=0.43}
\caption{Calculated neutrino fluxes from five Milagro hotspots,
assuming an $E^{-2}$ flux and gamma-ray cut-off at 300\,TeV. }
\label{fig2}
\end{figure}

Determining the flux of neutrinos from measurements of a pionic
gamma-ray spectrum is straightforward, as both are the decay products
of pions produced in proton-proton collisions. Here we calculate the
neutrino spectra using the method of \cite{apj:656:870}. It is
illustrated in Fig.~\ref{fig1}, comparing the gamma-ray spectrum from
H.E.S.S./MGRO\,J1908+06 to the calculated neutrino flux at Earth. As
the Milagro data extend to $\sim100$\,TeV without seeing a cut-off, we
take the gamma-ray cut-off at 300 TeV, corresponding to a proton
cut-off at energies of the order of the `knee'. The calculated
neutrino spectra from the five Milagro hotspots considered here are
shown in Fig.~\ref{fig2}, assuming an $E^{-2}$ spectrum normalized to
the Milagro measurement and also assuming a 300\,TeV gamma-ray
cut-off.

Earlier work on neutrino event rates from Milagro sources
\cite{pr:d76:067301,*pr:d75:083001,*pr:d76:123003} modeled the proton
spectrum in supernova remnants and investigated its effect on
gamma-ray and neutrino fluxes produced \emph{inside} the
accelerators. Given the evidence discussed in the previous section, we
assume in this work that the observed gamma rays are not produced
directly in the sources but in nearby molecular clouds resulting in
harder spectra which extend up to several 100\,TeV.

Neutrino telescopes detect the Cherenkov radiation from secondary
particles produced in the interactions of high-energy neutrinos in
highly transparent and well shielded deep water or ice with an array
of photomultipliers. They take advantage of the large cross section of
high-energy neutrinos and the long range of the muons produced. The
IceCube telescope \cite{app:26:155} is under construction and will
start taking data with a partial array of 2400 ten inch
photomultipliers positioned between 1500 and 2500\,m and deployed as
beads on 40 strings below the geographic South Pole. With the
completion of the detector by 2010--2011 the instrumented volume will
be doubled from 0.5 to $1\,{\rm km^{3}}$.

The event rate in a detector above a threshold energy $E_{\rm thresh}$
from a neutrino flux $dN_{\nu}/dE$ is given by
\begin{displaymath}
N_\mathrm{events} = T\int_{E_\mathrm{thresh}}A_{\mathrm eff}(E_{\nu})\,\frac{dN_{\nu}}{dE}\left(E_{\nu}\right)dE_{\nu},
\end{displaymath}
where the energy-dependent muon-neutrino effective area $A_{\rm
eff}(E_{\nu})$ is taken from \cite{proc:taup07:montaruli:1}. The
angular resolution is simulated as a function of the zenith angle
according to \cite{app:26:155} and lies between $0.7^\circ$ and
$0.8^\circ$ for the Milagro sources. The energy resolution is assumed
to be $\pm0.3$ in $\log(E_\nu)$ which seems achievable given the
superior performance of IceCube compared to AMANDA ($\sim 0.6$ in
$\log(E_\nu)$ taking \cite{proc:icrc07:muenich:1} and accounting for
the kinematics at the neutrino-muon vertex and the energy losses of
the muon on its way to the detector). The flux of atmospheric
neutrinos from the interactions of cosmic-ray protons n the Earth's
atmosphere, an irreducible background, is tabulated in
\cite{sovjnp:31:6:784} and gives a good parameterization of the AMANDA
measurements. Also, we assume no significant contribution from the
decay of charmed particles. We take the size of the search bin to be
the radius that gives $\sim70\%$ of the measured gamma-ray flux
assuming Gaussianity of the source emission and the angular
resolution. Table~\ref{tab1} lists the search bin radii.
Given a mean number of background events and a total number of
observed events (obtained using Poisson statistics for the sum of
signal and background taking into account 30\% signal reduction), we
calculate the probability (p value) that the observed number of events
is due to random fluctuations in the background. We define the
significance as the p value for which 50\% of experiments yield an
equal or lower p value.
\begin{table}
\begin{tabular}{ccccc}
  {\bf source } & {\bf \boldmath $r_\mathrm{bin}$ ($^\circ$)} &\hspace*{3mm}& 
  {\bf source } & {\bf \boldmath $r_\mathrm{bin}$ ($^\circ$)}\\ \hline
  MGRO J2019+37 & 1.4 &&  MGRO J2043+36 & 1.5 \\
  MGRO J1908+06 & 1.1 &&  MGRO J2032+37 &  1.5\\
  MGRO J2031+41 & 1.6 &&  (MGRO J1852+01 & 1.3)\\\hline
\end{tabular}
\caption{Angular radius of the IceCube search bin for each Milagro source.}
\label{tab1}
\end{table}

\begin{figure}
\epsfig{file=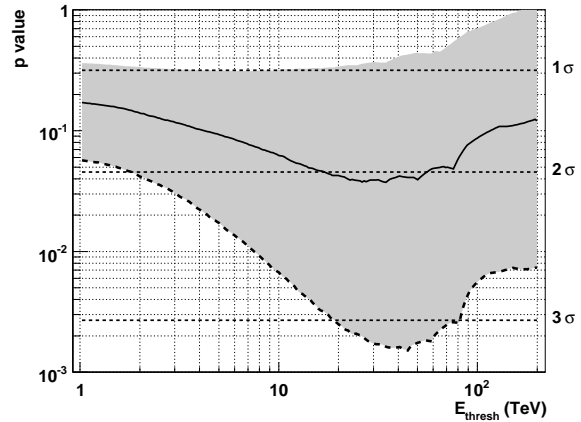, scale=0.43}
\caption{Significance of excess above background as a function of threshold
energy from H.E.S.S./MGRO\,J1908+06 after 10 years. The shaded area
represents the uncertainty in the H.E.S.S. $\gamma$-ray
measurements. The Milagro data points suggest the lower limit (dashed
line).}
\label{fig3}
\end{figure}

Figure \ref{fig3} shows the significance as a function of threshold
energy for MGRO\,J1908+06 after 10 years' data taking. Because of the
Milagro data points lying in the upper error range of the fitted gamma
spectrum (Fig.~\ref{fig1}), the Milagro measurements favor the
higher-significance range. As the significance of the excess will
likely be low even after 10 years, it may be necessary to use a
stacked search that will look for correlations between all five
Milagro sources of interest and the IceCube sky map simultaneously.

\begin{figure}
\epsfig{file=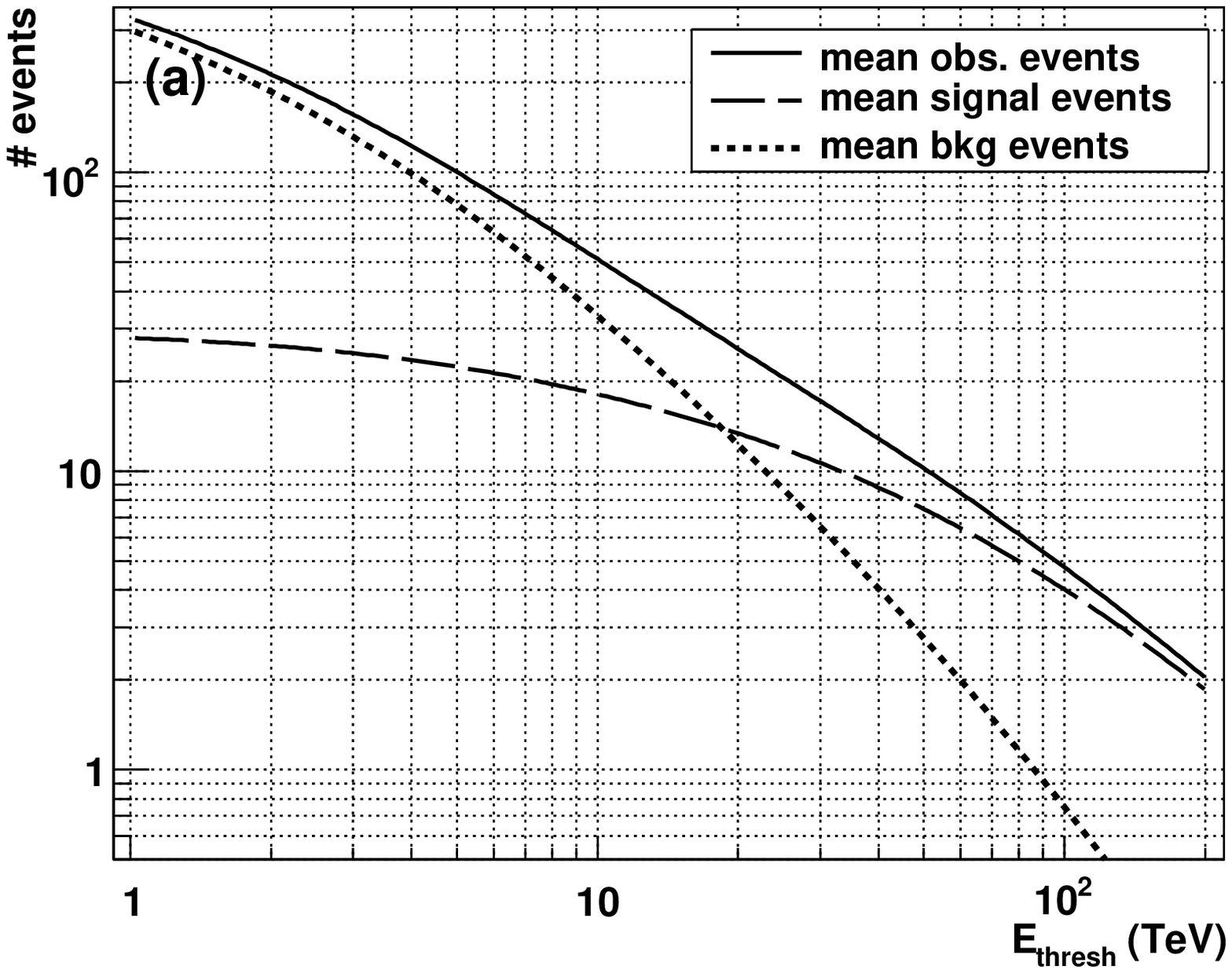, scale=0.43}
\epsfig{file=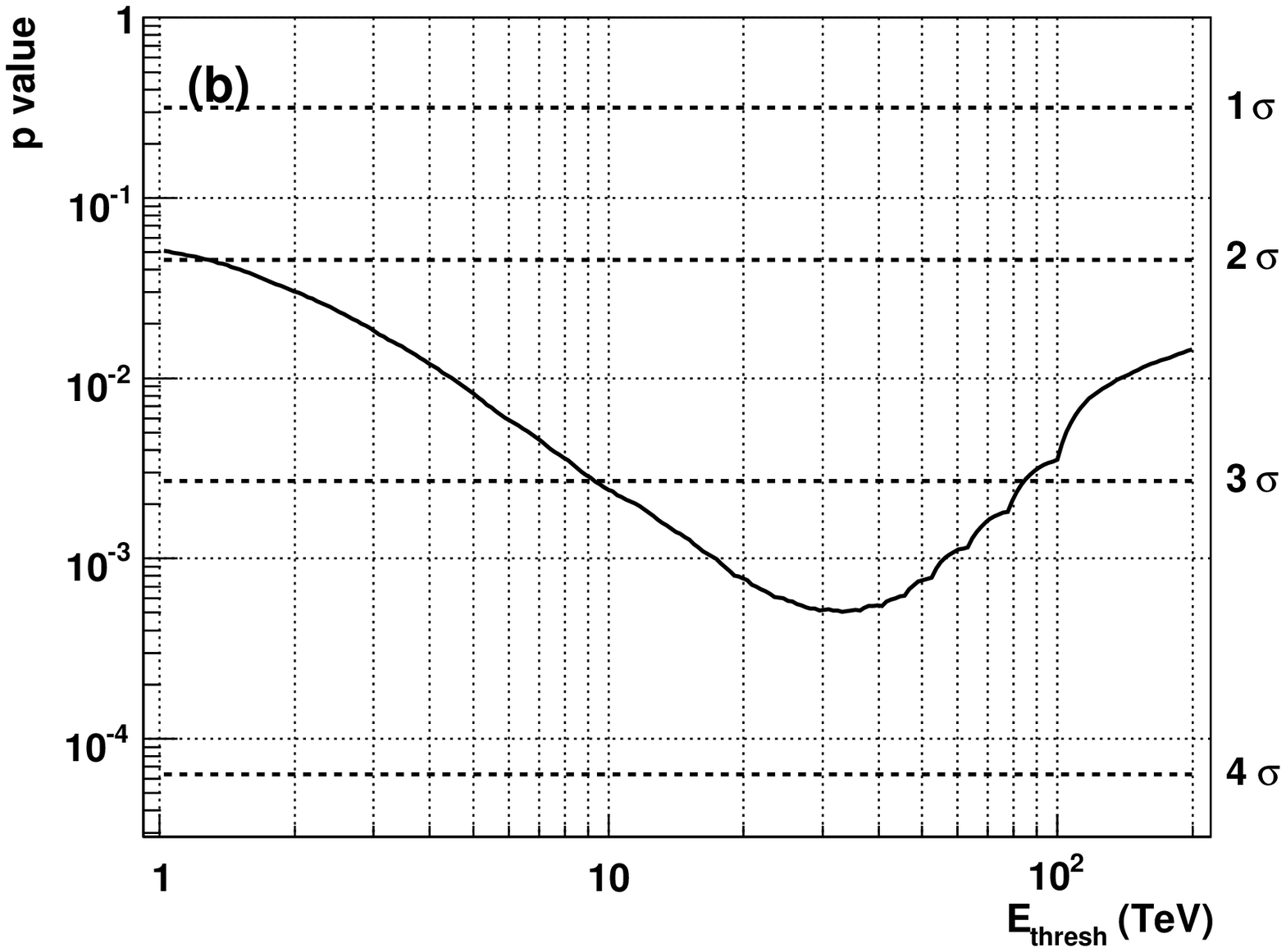, scale=0.43}
\caption{(a) Mean number of neutrinos from the Milagro hotspots as a
function of energy threshold (dashed line) compared to the background
(dotted line) and total mean (solid line) number of events from the
search bins in 10 years. (b) Corresponding significance of observed
excess from the Milagro hotspots.}
\label{fig4}
\end{figure}

\begin{figure}
\epsfig{file=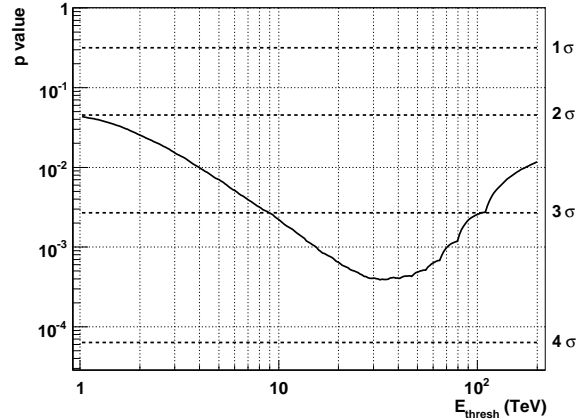, scale=0.43}
\caption{Significance of observed excess from the Milagro hotspots
including the potential hotspot MGRO\,J1852+01 after 5 years
of observation time.}
\label{fig5}
\end{figure}

Figure~\ref{fig4}a shows the mean number of signal events in IceCube
in 10 years from the five Milagro sources (excluding MGRO\,1852+01) as
a function of energy threshold together with the mean total number of
events. The significance of the correlation of this catalog with the
IceCube sky map after 10 years' time is given by Fig.~\ref{fig4}b. If
the potential hotspot MGRO\,J1852+01 turns out to be real the same
significance would be reached after only 5 years' observation time
(Fig.~\ref{fig5}).  The figures make clear that the best prospect for
detecting these sources is to focus on events above several tens of
TeV, where the atmospheric background is very low but there are still
sufficient signal events left. Then, a detection of these sources
after several years is possible.

The results we obtain are conservative in several ways. The quoted
angular resolution is based on simulations assuming AMANDA technology
and reconstruction methods. A not unrealistic increase in the
resolution from $0.7^\circ$--$0.8^\circ$ to $0.5^\circ$ improves the
significance in Fig.~\ref{fig4}b to $4\sigma$. Also, the assumed
photon fluxes at the sources might be higher due to absorption in the
Galactic photon background (especially in the Cygnus region), or, in
the case of MGRO\,J1908+06, due to currently ambiguous measurements
(see Fig.~\ref{fig1}). An overall flux increase by 20\% (50\%) boosts
the significance in Fig.~\ref{fig4}b to $4\sigma$ ($5\sigma$). Even
reducing the energy resolution to $\pm 0.5$ in $\log(E_\nu)$ still
results in a significance of better than $3
\sigma$. Furthermore, the use of methods such as unbinned searches
beyond the simple binned method considered here will increase
IceCube's sensitivity \cite{arxiv:0801.1604}.

\section{Conclusions}
Apart from the Crab nebula, Milagro has clearly identified three
sources of high-energy gamma-ray emission in their skymap. Two of
these sources (MGRO\,J2019+37 and MGRO\,J2031+41) are located in the
Cygnus region and one (MGRO\,J1908+06) closer to the Galactic
center. Furthermore, the Cygnus region contains two candidate hotspots
(MGRO\,J2043+36 and MGRO\,J2032+37). Another potential hotspot
(MGRO\,J1852+01) falls currently below the significance threshold set
by the Milagro collaboration. Combining these measurements with
measurements from Cherenkov telescopes shows that several of these
sources have unusual hard spectra consistent with $E^{-2}$, where in
one case the spectrum seems to extend up to 100\,TeV without
indication of a cut-off.

Sources producing such hard spectra extending up to 100\,TeV and more
are required to explain the existence of the `knee' in the cosmic-ray
spectrum around 3\,PeV, with young supernova remnants being the best
candidates. However, their observation is difficult as these
high-energy photons are produced inside the accelerator only within
the first few hundred years. Indeed, all current flux calculations for
the Milagro sources assume a gamma-ray production scenario inside the
acceleration region and in most scenarios predict photon spectra with
cut-offs below 100\,TeV.

In our paper we adopt the novel idea that a cosmic-ray source can
produce a hard gamma-ray spectrum up to high energies over a much
longer time of several thousand years if the gamma-rays are produced
\emph{outside} the acceleration region in the interaction of the
source's Pevatron beam with a nearby molecular cloud
\cite{apjl:665:l131}. Assuming an $E^{-2}$ spectrum with a cut-off at
300\,TeV (consistent with a proton cut-off at the `knee') we
demonstrate that IceCube will be able to see these sources after
several years of observation. For the significance calculations we use
results from a detailed detector simulation and take into account the
energy resolution and zenith-angle-dependent angular resolution. The
former is especially important as we demonstrate that the highest
sensitivity is obtained by specializing to events with energies above
30\,TeV. Similar studies have so far neglected energy migration
effects.

With a visibility of 50\%, the location of MGRO\,J1908+06 puts it
within reach of a future ${\rm km^{3}}$-scale Mediterranean neutrino
telescope. However, in the southern hemisphere no Pevatron candidates
have been discovered so far, perhaps because no all-sky instrument
like Milagro is currently operational in that hemisphere.  While a
high-resolution pointed telescope could resolve what previously
appeared to be a diffuse source into its individual parts (supernova
remnants and molecular clouds), it is possible that the high density
of ambient matter in star-forming regions prevents individual sources
from dominating the off-source flux to give sufficient statistical
significance for a pointing telescope. In this case a Milagro-like
telescope with a broader field of view such that background
measurements are truly ``off-source'' would be needed. In this context
it is interesting to note that IceCube may be able to detect
gamma-rays from the southern sky
\cite{future:halzenetal:1} and therefore be used to search for
southern Pevatrons over a broad range similar to Milagro's.
\begin{acknowledgments}
{\footnotesize This research was supported in part by the National
Science Foundation under Grant No.~OPP-0236449, in part by the
U.S.~Department of Energy under Grant No.~DE-FG02-95ER40896, and in
part by the University of Wisconsin Research Committee with funds
granted by the Wisconsin Alumni Research Foundation. A.K. acknowledges
support by the EU Marie Curie OIF program.}
\end{acknowledgments}

\ifx\mcitethebibliography\mciteundefinedmacro
\PackageError{unsrtM.bst}{mciteplus.sty has not been loaded}
{This bibstyle requires the use of the mciteplus package.}\fi

\end{document}